\documentclass[12pt]{article}
\usepackage{epsf}
\makeatletter
\@addtoreset{equation}{section}

\renewcommand{\thefootnote}{\fnsymbol{footnote}}
\makeatother

\newcommand{\Lm}{\Lambda}

\newcommand{\Acl}{A_{\rm cl}}

\newcommand{\am}{{\rm am}}
\newcommand{\sn}{{\rm sn}}
\newcommand{\cn}{{\rm cn}}
\newcommand{\dn}{{\rm dn}}
\newcommand{\gef}[1]{g_{{\rm eff}#1}}

\newcommand{\AR}{A_{\rm R}}
\newcommand{\AI}{A_{\rm I}}
\newcommand{\ar}[1]{a_{{\rm R},#1}}
\newcommand{\ai}[1]{a_{{\rm I},#1}}
\newcommand{\bR}[1]{b_{{\rm R},#1}}
\newcommand{\bI}[1]{b_{{\rm I},#1}}
\newcommand{\mR}[1]{m_{{\rm R},#1}}
\newcommand{\mI}[1]{m_{{\rm I},#1}}
\newcommand{\f}[2]{f^{(#1)}_{#2}}
\newcommand{\ps}[2]{\psi^{(#1)}_{#2}}
\newcommand{\CR}[1]{C_{{\rm R},#1}}

\newcommand{\cL}{{\cal L}}
\newcommand{\del}{\partial}

\newcommand{\dls}{\partial\hspace{-6.5pt}/}

\newcommand\be{\begin{equation}}
\newcommand\ee{\end{equation}}
\newcommand\bea{\begin{eqnarray}}
\newcommand\eea{\end{eqnarray}}

\begin{document}

\thispagestyle{empty}
%\begin{titlepage}
%
\begin{flushright}
TIT/HEP--473 \\
UT-984\\
{\tt hep-th/yymmnn} \\
December, 2001 \\
\end{flushright}
\vspace{3mm}
\begin{center}
{\Large
{\bf  SUSY Breaking by stable non-BPS configurations
 }} 
%\\[18mm]
%\lineskip .75em
\vskip 1.5cm

%\normalsize

{\bf Nobuhito Maru~$^{a}$}
\footnote{\it  e-mail address: 
maru@hep-th.phys.s.u-tokyo.ac.jp},  
{\bf 
Norisuke Sakai~$^{b}$}
\footnote{Speaker at the conference, 
{\it  e-mail address: nsakai@th.phys.titech.ac.jp} 
}
{\bf 
Yutaka Sakamura~$^{b}$}
\footnote{\it  e-mail address: sakamura@th.phys.titech.ac.jp} 
%\\
%\bigskip
~and~~ {\bf 
Ryo Sugisaka~$^{b}$}
\footnote{\it  e-mail address: sugisaka@th.phys.titech.ac.jp}

\vskip 1.5em

{ \it   $^{a}$Department of Physics, University of Tokyo 
113-0033, JAPAN \\
and \\
$^{b}$Department of Physics, Tokyo Institute of Technology 
\\
Tokyo 152-8551, JAPAN  }
\vspace{10mm}
{\bf Abstract}\\[5mm]
{\parbox{14cm}{\hspace{5mm}
%%%%%%%%%%%%%%%%%%%%%%%%%%%%%%%%%%%%%%%%%%%%%%%%%%
%%%%%%%%%%
A simple mechanism for SUSY breaking is proposed due to the 
coexistence of BPS domain walls. 
It requires no messenger fields 
nor complicated SUSY breaking sector on any of the walls. 
We assumed that our world is on a BPS domain wall and that the other 
BPS wall breaks the SUSY preserved by our wall. 
We obtain an ${\cal N}=1$ model in four dimensions which 
admits an exact solution of a stable non-BPS configuration of two walls. 
The stability is assured by a topological quantum number associated 
with the winding on the field space of the topology of $S^1$. 
We propose that the overlap of 
the wave functions of the Nambu-Goldstone fermion and those of physical fields 
provides a practical method to evaluate SUSY breaking mass splitting 
on our wall thanks to a low-energy theorem. 
This is based on our recent works hep-th/0009023 and hep-th/0107204. 
%%%%%%%%%%%%%%%%%%%%%%%%%%%%%%%%%%%%%%%%%%%%%%%%%%
%%%%%%%%%%
}}
\end{center}
\vfill
\newpage
\setcounter{page}{1}
\setcounter{footnote}{0}
\renewcommand{\thefootnote}{\arabic{footnote}}

%%%%%%%%%%%%%%%%%%%%%%%%%%%%%%%%%%%%%%
\section{Introduction}\label{INTRO}
%%%%%%%%%%%%%%%%%%%%%%%%%%%%%%%%%%%%%%
Supersymmetry (SUSY) provides the most realistic models to 
solve the  hierarchy problem in unified theories \cite{DGSW}. 
One of the most important issues in model building of SUSY 
unified theories  has been for some years how to understand 
the SUSY breaking in our observable world. 
Many models of SUSY breaking uses some kind of mediation 
of the SUSY breaking 
from the hidden sector to our observable sector. 

Recently the ``Brane World" scenario has become quite popular 
where  our four-dimensional spacetime is realized on the wall 
embedded in a higher dimensional spacetime \cite{LED,RS}. 
In order to discuss the stability of such a wall, it is often 
useful to consider SUSY theories as the fundamental theory. 
Moreover, SUSY theories in higher dimensions are a natural 
possibility in string theories. 
These SUSY theories in higher dimensions have $8$ or more 
supercharges, which should be broken partially 
if we want to have a phenomenologically viable SUSY unified model 
in four dimensions. 
Such a partial breaking of SUSY is nicely obtained by topological 
defects \cite{WittenOlive}. 
Domain walls or other topological defects preserving 
part of the original SUSY in the fundamental theory 
are called the BPS states in SUSY theories. 
Walls have co-dimension one and typically 
 preserve half of the original SUSY, which are called 
 $1/2$ BPS states  \cite{CGR,DW,KSS}. 
 Junctions of walls have co-dimension two and typically 
 preserve 
 a quarter of the original SUSY \cite{AbrahamTownsend,DWJ}. 

The new possibility offered by the brane world 
scenario stimulated  studies of SUSY breaking. 
Recently we have proposed a simple mechanism of SUSY 
breaking due to the 
coexistence of different kinds of BPS domain walls and 
proposed an efficient 
method to evaluate the SUSY breaking parameters such as 
the boson-fermion mass-splitting 
by means of overlap of wave functions involving the 
Nambu-Goldstone (NG) fermion \cite{MSSS}--\cite{MSSStw}. 
We have exemplified these points by taking a toy model in 
four dimensions, 
which allows an exact solution of coexisting walls with a 
three-dimensional effective theory \cite{MSSS}. 
Although the first model is only meta-stable, we were able to 
show approximate 
evaluation of the overlap allows us to determine 
the mass-splitting reliably. 
More recently, we have constructed a stable non-BPS configuration of 
two walls in an ${\cal N}=1$ supersymmetric model in four 
dimensions to demonstrate our idea of SUSY 
breaking due to 
the coexistence of BPS walls. 
We have also extended our analysis 
to more realistic case of four-dimensional effective theories 
and examined the consequences of our mechanism in detail \cite{MSSS2}. 

Our proposal for a SUSY breaking mechanism requires no 
messenger fields, 
nor  complicated SUSY breaking sector on any of the walls. 
We assume that our world is on a wall and SUSY is broken 
only by the 
coexistence of another wall with some distance from our wall. 
The NG fermion is localized on the distant 
wall and its overlap with the wave functions of physical fields 
on our wall gives the boson-fermion mass-splitting of 
physical fields on our 
wall thanks to a low-energy theorem \cite{lee-wu}. 

The purpose of this paper is to illustrate our idea of SUSY 
breaking due to 
the coexistence of BPS walls by taking a simple soluble 
model with a stable 
non-BPS configuration of two walls and to extend our analysis 
to more realistic case of four-dimensional effective theories. 
We work out how various soft SUSY breaking terms 
can arise in our framework. 
Phenomenological implications are briefly discussed. 
We also find that effective SUSY breaking scale observed on 
our wall becomes 
exponentially small as the distance between two walls grows. 
The NG fermion is localized on the distant 
wall and its overlap with the wave functions of physical fields 
on our wall gives the boson-fermion mass-splitting of 
physical fields on our 
wall thanks to a low-energy theorem. 
We have proposed that this overlap provides a practical 
method to evaluate 
the mass-splitting in models with SUSY breaking due to the 
coexisting walls. 

%%%%%%%%%%%%%%%%%%%%%%%%%%%%%%%%%%%%%%%%%%%%%%%%%%
%%%
\section{BPS equation and topological quantum number 
} 
\label{BPS-winding}
%%%%%%%%%%%%%%%%%%%%%%%%%%%%%%%%%%%%%%%%%%%%%%%%%%
%%%
Let us illustrate the BPS equation and topological quantum number 
for ${1 \over 2}$-BPS state in terms of a simple model 
with a single chiral scalar field $\Phi'=(A',\psi',F')$ and a 
superpotential 
${\cal W}=%\Lambda^{2}
\Phi'-{1 \over 3}%\frac{g}{3}
\Phi'^{3}$. 
After eliminating the auxiliary field $F$ the bosonic part of 
the Lagrangian becomes  
\begin{equation}
{\cal L}=
-\partial_\mu A'^{*} \partial^\mu A' 
- \left| 1 - A'^2 \right|. 
\end{equation}
We have absorbed possible constants into the normalization of field 
and coordinates for simplicity. 
The model has two SUSY vacua at $A'=\pm 1$. 
The supertransformation of the fermion $\psi'$ is given by 
\begin{equation}
\delta \psi'=
i\sqrt2 \sigma^\mu \bar \epsilon \partial_\mu A' +\sqrt2 \epsilon F'. 
\end{equation}
If we choose $A'$ to depend only on one coordinate, say, $x^2=y$, 
and choose $\epsilon=i\sigma^2 \bar \epsilon$, 
the half of supersymmetry is conserved by the configuration 
satisfying the BPS equation 
\begin{equation}
{d A' \over dy}=  1 - A'^2. 
\label{eq:BPSeq_cubic}
\end{equation}
It admits a wall solution connecting the SUSY vacuum $-1$ 
at $y=-\infty$ to another SUSY vacuum  $+1$ 
at $y=\infty$   
\begin{equation}
A'^{(1)}_{\rm cl} (y)=  \tanh (y-y_1), 
\label{eq:BPSsol_cubic}
\end{equation}
where $y_1$ denotes the position of the wall. 
Orthogonal linear combination of supercharges are conserved if the 
anti-BPS equation is satisfied 
\begin{equation}
{d A' \over dy}= - (1 - A'^2), 
\end{equation}
which admits a wall solution connecting the SUSY vacuum $+1$ 
at $y=-\infty$ to another SUSY vacuum  $-1$ 
at $y=\infty$   
\begin{equation}
A'^{(2)}_{\rm cl} (y)= - \tanh (y-y_2), 
\end{equation}
where $y_2$ denotes the position of the wall. 
If we combine these two solutions, we obtain a wall anti-wall 
configuration. 
In fact we have found exact solution of the equation of motion 
which is a non-BPS state and gives an example of the SUSY breaking 
due to the coexistence of BPS and anti-BPS walls \cite{MSSS}. 
The wall anti-wall configuration is unstable due to the annihilation 
into vacuum. 
It is desirable to have a model with stable but non-BPS two wall 
configuration. 

We have found a way to give the topological quantum number. 
We shall give a topology of $S^1$ to field space so that we can have 
a notion of winding from a compactified base space which is also $S^1$. 
To achieve that goal, we change field variable $A'$ 
into a periodic variable $A$ 
\begin{equation}
A' = \sin A, \qquad \Phi'=\sin \Phi.
\end{equation}
Then the SUSY vacua occurs at $A=\pi \left(n+{1 \over 2}\right)$ 
with the periodicity $A=A+2\pi$. 
The BPS equation (\ref{eq:BPSeq_cubic}) becomes 
\begin{equation}
{d A \over dy}=  \cos A. 
\label{eq:BPSeq_periodic}
\end{equation}
The BPS solution (\ref{eq:BPSsol_cubic}) is mapped into 
a solution of this transformed BPS Eq.(\ref{eq:BPSeq_periodic}) 
\begin{equation}
\sin A^{(1)}_{\rm cl} (y)= \tanh (y-y_1) 
\end{equation}
connecting the SUSY vacuum $A=-\pi/2$ at $y=-\infty$ to $A=\pi/2$ 
at $y=\infty$. 
The solution of the anti-BPS equation 
connecting the SUSY vacuum $A=\pi/2$ at $y=-\infty$ to $A=3\pi/2$ 
at $y=\infty$. can also be obtained 
\begin{equation}
\sin A^{(2)}_{\rm cl} (y)= - \tanh (y-y_2) .
\end{equation}
Now we can see these solutions can be smoothly connected in the field space 
since the field $A=\pi/2$ at right end point of the BPS wall 
is the same as the field at the left end point of the anti-BPS wall. 
This suggests that we may have a non-BPS solution of two wall configuration 
which is non-BPS. 
Indeed we found that a simple model with minimal kinetic term provides 
the BPS equation (\ref{eq:BPSeq_cubic}) and 
 that there is an exact solution for the non-BPS configuration 
of two walls which winds around the field space $A$ once \cite{MSSS2} 
\footnote{
Since the $1/2$-BPS solution is intrinsically real, we can interprete 
the nonlinearity of the right-hand side of the BPS equation in two ways: 
Either as the derivative of a superpotential (our model), or 
as the inverse of the nontrivial K\"ahler metric in a nonlinear sigma model 
as in \cite{NNS}. 
Here we choose a simpler possibility. 
}. 

%%%%%%%%%%%%%%%%%%%%%%%%%%%%%%%%%%%%%%%%%%%%%%%%%%
%%%
\section{Stable non-BPS configuration of two walls
%SUSY breaking by the coexistence of walls
} 
\label{SUSY-br-coexist}
%%%%%%%%%%%%%%%%%%%%%%%%%%%%%%%%%%%%%%%%%%%%%%%%%%

In order to illustrate our basic ideas, 
 we consider three dimensional domain walls in four-dimensional spacetime. 
Our model reads 
\begin{eqnarray}
 &\!\!\!\cL&\!\!\!=\bar{\Phi}\Phi |_{\theta^{2}\bar{\theta}^{2}}
  +W(\Phi)|_{\theta^{2}}+{\rm h.c.},  \label{Logn} 
\qquad 
W(\Phi)
=\frac{\Lm^{3}}{g^{2}}\sin\left(\frac{g}{\Lm}\Phi\right). 
%\nonumber
\end{eqnarray}
We have introduced a scale parameter $\Lm$ with a mass-dimension one and a 
 dimensionless coupling constant $g$, both of which are real positive.
Choosing $y=X^{2}$ as the extra dimension 
and compactify it on $S^{1}$ of radius $R$.
Other coordinates are denoted as $x^{m}$ ($m=0,1,3$), {\em i.e.}, 
$X^{\mu}=(x^{m},y)$. 
The bosonic part of the model is 
\be
 \cL_{\rm 
bosonic}=-\del^{\mu}A^{\ast}\del_{\mu}A-\frac{\Lm^{4}}{g^{2}}
 \left|\cos\left(\frac{g}{\Lm}A\right)\right|^{2}.
\ee
The target space of the scalar field $A$ has a topology of a 
cylinder. 
This model has two vacua at $A=\pm\pi\Lm/(2g)$, both lie 
on the real axis.

In the limit $R\to\infty$, 
we have a BPS domain wall solution 
\be
 \sin {g \over \Lambda} \Acl^{(1)}(y)=
 \tanh \left(\Lm(y-y_{1})\right),
 \label{eq:first_wall}
\ee
which interpolates the vacuum at $A=-\pi\Lm/(2g)$ to that 
at 
$A=\pi\Lm/(2g)$ as $y$ increases from $y=-\infty$ to 
$y=\infty$ and conserves the real two componet SUSY charge 
$Q_\alpha^{(1)}$ which can be regarded as supercharges in 
three dimensions. 
We have also an anti-BPS wall solution 
\be
 \sin {g \over \Lambda} \Acl^{(2)}(y)=
 - \tanh \left(\Lm(y-y_{2})\right),
 \label{eq:second_wall}
\ee
which interpolates the vacuum at $A=\pi\Lm/(2g)$ to that at 
$A=3\pi\Lm/(2g)=-\pi\Lm/(2g)$ and preserves another 
real two component supercharge $Q_\alpha^{(2)}$. 
Here $y_{1}$ and $y_{2}$ are integration constants and 
represent 
the location of the walls along the extra dimension. 
The four-dimensional supercharge $Q_{\alpha}$ is a sum of these two 
supercharges 
$%\be
 Q_{\alpha}=\frac{1}{\sqrt{2}}(Q^{(1)}_{\alpha}+iQ^{(2)}_{\alpha})
$. %\ee
Each wall breaks a half of the bulk supersymmetry and 
all of the bulk supersymmetry will be broken if these 
walls coexist.

{}For this model, we have found an exact solution of the non-BPS two wall 
configuration which is stable due to the winding number: $\pi(S^1)=Z$. 
Such a configuration should be a solution of the equation of 
motion, 
\be
 \del^{\mu}\del_{\mu}A+\frac{\Lm^{3}}{g}\sin\left(\frac{g}{\Lm}
A^{\ast}\right)
 \cos\left(\frac{g}{\Lm}A\right)=0. \label{EOM1}
\ee
A general real static solution of Eq.(\ref{EOM1}) 
that depends 
only on $y$ is found to be 
\be
 \Acl(y)=\frac{\Lm}{g}\am\left(\frac{\Lm}{k}(y-y_{0}),k\right), 
 \label{A_classical}
\ee
where $k$ and $y_{0}$ are real parameters and the function 
$\am(u,k)$ 
denotes the amplitude function, which is defined 
as an inverse function of 
$%\be
 u(\varphi)=\int_{0}^{\varphi}\frac{{\rm d}\theta}
 {\sqrt{1-k^{2}\sin^{2}\theta}}
$. %\ee
If $k<1$, the solution $\Acl(y)$ is a monotonically increasing 
function with 
\be
 \Acl\left(y+{4kK(k) \over \Lambda}\right)= \Acl(y)+ 2 
\pi{\Lambda \over g}.
\ee
%%%%%%%%%%%%%%%%%%%%%%%% figure 
%%%%%%%%%%%%%%%%%%%%%%%%%%%%%
\begin{figure}[t]
 \leavevmode
 \epsfysize=8cm
 \centerline{\epsfbox{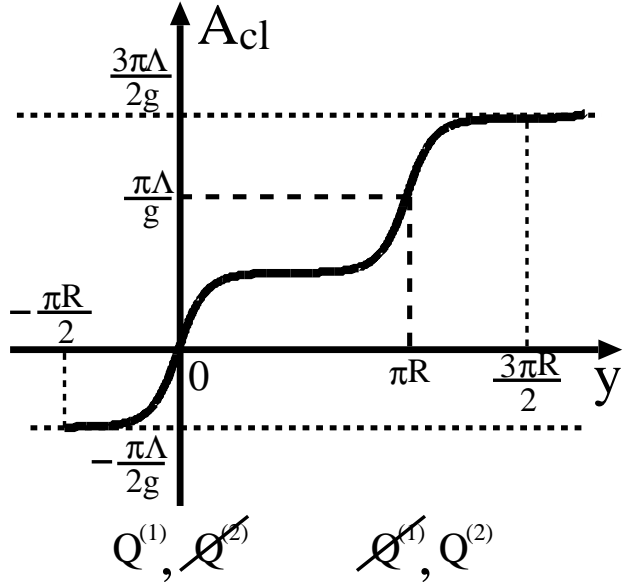}}
 \caption{The profile of the classical solution $\Acl(y)$. 
 The dotted lines $A=-\pi\Lm/(2g)$ and $A=3\pi\Lm/(2g)$ 
are identified.
 }
 \label{profile-Acl}
\end{figure}
%%%%%%%%%%%%%%%%%%%%%%%%%%%%%%%%%%%%%%%%%%%%%%%%%%
%%%%%%%%%%%
This is the solution that we want.
Since the field $A$ is an angular variable 
$A=A+2\pi\Lambda/g$, 
we can choose the compactified radius 
$2\pi R=4kK(k)/\Lm$ so that the classical field configuration 
$\Acl(y)$ contains two walls and becomes 
periodic modulo $2\pi\Lambda/g$. 
We shall take $y_{0}=0$ to locate one of the walls at $y=0$. 
Then we find that the other wall is located at the anti-podal 
point 
 $y=\pi R$ of the compactified circle. 
We have computed the energy of a superposition of the first 
wall 
$\Acl^{(1)}(y)$ located at $y=y_1$ in Eq.(\ref{eq:first_wall}) 
 and the second wall $ \Acl^{(2)}(y)$ located at $y=y_2$ 
in Eq.(\ref{eq:second_wall}). 
This energy can be regarded as a potential between two walls 
in the adiabatic 
approximation and has a peak at $|y_1-y_2|=0$ implying that 
two walls 
experience a repulsion. 
This is in contrast to a BPS configuration of two walls which 
should exert 
no force between them. 
Thus we can explain that the second wall is settled at the 
anti-podal point $y=\pi R$ in our stable non-BPS 
configuration because of 
the repulsive force between two walls. 
Since the repulsive force forces the other wall to oscillate around 
the anti-podal point when a small fluctuation is added, 
we have a physical reason to obtain a stable spectrum without any tachyon. 

In the limit of $R\to\infty$, {\em i.e.}, $k\to 1$, $\Acl(y)$ 
approaches near $y=0$ to 
the BPS configuration $\Acl^{(1)}(y)$ with $y_1=0$ which 
preserves $Q^{(1)}$, 
and near $y=\pi R$ to $\Acl^{(2)}(y)$ with $y_2=\pi R$ which preserves 
$Q^{(2)}$. 
The profile of the classical solution $\Acl(y)$ is shown 
in Fig.\ref{profile-Acl}. 
We will refer to the wall at $y=0$ as ``our wall'' and the wall at 
$y=\pi R$ as ``the other wall''.

\section{Mode expansion and effective Lagrangian}
The fluctuation fields around the background $\Acl(y)$ can be expanded into 
modes 
\bea
 A(X)&\!\!=&\!\!\Acl(y)+\frac{1}{\sqrt{2}}(\AR(X)+i\AI(X)), 
 \nonumber\\
 \Psi_{\alpha}(X)&\!\!=&\!\!\frac{1}{\sqrt{2}}(\Psi_{\alpha}^{(1)}(
X)
 +i\Psi_{\alpha}^{(2)}(X)). \label{fluc_fields}
\eea
The four-dimensional fluctuation fields can be expanded as 
\be
 \AR(X)=\sum_{p}\bR{p}(y)\ar{p}(x),\;\;\;
 \AI(X)=\sum_{p}\bI{p}(y)\ai{p}(x), 
\label{eq:boson_mode_decomp}
\ee
\be
 \Psi^{(1)}(X)=\sum_{p}\f{1}{p}(y)\ps{1}{p}(x),\;\;\;
 \Psi^{(2)}(X)=\sum_{p}\f{2}{p}(y)\ps{2}{p}(x). 
 \label{eq:fermion_mode_decomp}
\ee

%%%%%%%%%%%%%%%%%%%%%%%%% figure 
%%%%%%%%%%%%%%%%%%%%%%%%%%%
\begin{figure}[t]
 \leavevmode
 \epsfysize=6cm
 \centerline{\epsfbox{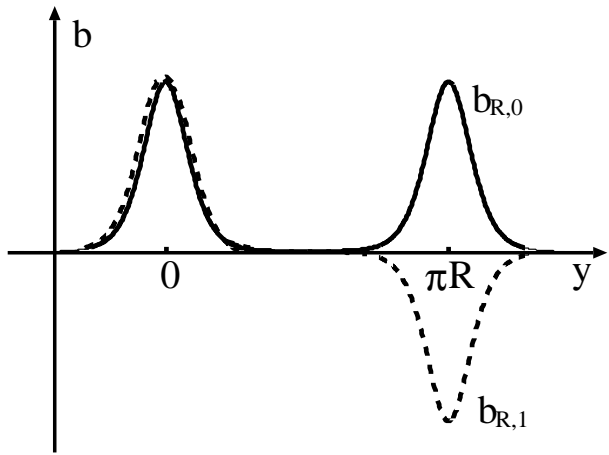}}
 \caption{The mode functions for the bosonic modes $\ar{0}$ 
 and $\ar{1}$. The solid line represents the profile of 
$\bR{0}(y)$ and 
 the dashed line is that of $\bR{1}(y)$.
 }
 \label{boson-mode}
\end{figure}
%%%%%%%%%%%%%%%%%%%%%%%%%%%%%%%%%%%%%%%%%%%%%%%%%%
%%%%%%%%%%%
Exact mode functions and mass-eigenvalues can be found for 
several light modes 
of $\bR{p}(y)$, 
\bea
 \bR{0}(y)&\!\!=&\!\!\CR{0}\dn\left(\frac{\Lm y}{k},k\right), 
 \;\;\; \mR{0}^{2}=0, \nonumber \\
 \bR{1}(y)&\!\!=&\!\!\CR{1}\cn\left(\frac{\Lm y}{k},k\right), \;
\;\; 
 \mR{1}^{2}=\frac{1-k^{2}}{k^{2}}\Lm^{2}, \nonumber \\
 \bR{2}(y)&\!\!=&\!\!\CR{2}\sn\left(\frac{\Lm y}{k},k\right), \;
\;\; 
 \mR{2}^{2}=\frac{\Lm^{2}}{k^{2}}, \label{bosonR_mode_fnc}
\eea
where functions $\dn(u,k)$, $\cn(u,k)$, $\sn(u,k)$ are the 
Jacobi's 
elliptic functions and $\CR{p}$ are normalization factors. 
{}For $\bI{p}(y)$, we can find all the eigenmodes 
\be
 \bI{p}(y)=\frac{1}{\sqrt{2\pi R}}{\rm e}^{i{p \over R}y}
% \sin\left(\frac{p}{R}(y-\alpha)\right)
,\;\;\; 
 \mI{p}^{2}=\Lm^{2}+\frac{p^{2}}{R^{2}},\;\;\;
 ( p \in {\bf Z} 
 %=0,1,2,\cdots
 ).
\ee
The massless field $\ar{0}(x)$ is the Nambu-Goldstone (NG) 
boson 
for the breaking of the translational invariance in the extra 
dimension.
The first massive field $\ar{1}(x)$ corresponds to the 
oscillation of the 
background wall around the anti-podal equilibrium point and 
hence becomes 
massless in the limit of $R\rightarrow \infty$. 
All the other bosonic fields remain massive in that limit. 

{}For fermions, only zero modes are known explicitly,
\be
 \f{1}{0}(y)=C_{0}\left\{\dn\left(\frac{\Lm y}{k},k\right)
 +k\cn\left(\frac{\Lm y}{k},k\right)\right\}, \;\;\;
 \f{2}{0}(y)=C_{0}\left\{\dn\left(\frac{\Lm y}{k},k\right)
 -k\cn\left(\frac{\Lm y}{k},k\right)\right\}, 
\label{fermion_mode_fnc}
\ee
where $C_{0}$ is a normalization factor.
These fermionic zero modes are the NG fermions for the 
breaking of 
$Q^{(1)}$-SUSY and $Q^{(2)}$-SUSY, respectively. 

%%%%%%%%%%%%%%%%%%%%%%%%% figure 
%%%%%%%%%%%%%%%%%%%%%%%%%%%
\begin{figure}[t]
 \epsfysize=5cm
 \centerline{\epsfbox{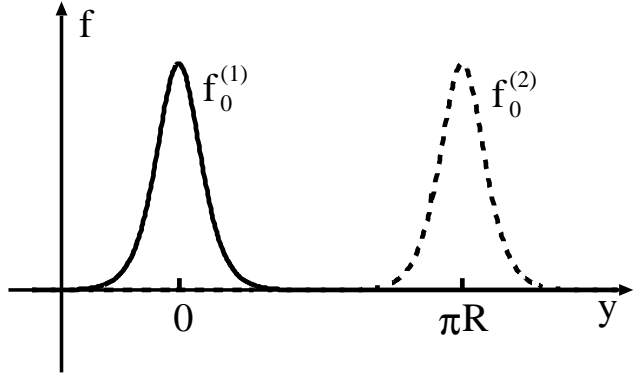}}
 \caption{The mode functions for fermionic zero-modes 
$\ps{1}{0}$ and
 $\ps{2}{0}$. The solid line represents the profile of $\f{1}{0}(y)
$ and 
 the dashed line is that of $\f{2}{0}(y)$.
 }
 \label{fermion-mode}
\end{figure}
%%%%%%%%%%%%%%%%%%%%%%%%%%%%%%%%%%%%%%%%%%%%%%%%%%
%%%%%%%%%%%

Thus there are four fields which are massless or become 
massless in the limit 
of $R\rightarrow \infty$: 
$\ar{0}(x)$, $\ar{1}(x)$, $\ps{1}{0}(x)$ and $\ps{2}{0}(x)$.
The profiles of their mode functions are shown in 
Fig.\ref{boson-mode} and 
Fig.\ref{fermion-mode}.
Other fields are heavier and have masses of the order of 
$\Lm$. 
We will concentrate ourselves on 
the breaking of the $Q^{(1)}$-SUSY, which is approximately 
preserved 
by our wall at $y=0$.
So we call the field $\ps{2}{0}(x)$ the NG fermion. 

We can obtain a three-dimensional effective Lagrangian 
by substituting the mode-expanded fields 
into the Lagrangian 
(\ref{Logn}), 
and carrying out an integration over $y$ 
\begin{eqnarray}
 \cL^{(3)}&\!\!\!=&\!\!\!
 -V_{0}-\frac{1}{2}\del^{m}\ar{0}\del_{m}\ar{0}
 -\frac{1}{2}\del^{m}\ar{1}\del_{m}\ar{1}-\frac{i}{2}\ps{1}{0}
 \dls\ps{1}{0}
 -\frac{i}{2}\ps{2}{0}\dls\ps{2}{0} \nonumber \\
 &\!\!\!&\!\!\!
 -\frac{1}{2}\mR{1}^{2}\ar{1}^{2}+\gef{}\ar{1}\ps{1}{0}\ps{2}{0}
 +\cdots, 
 \label{effthry}
\end{eqnarray}
where $\dls\equiv\gamma^{m}_{(3)}\del_{m}$ and 
an abbreviation denotes terms involving heavier fields and 
higher-dimensional terms. 
Here 
 $\gamma$-matrices in three dimensions are defined by 
 $\left(\gamma^{m}_{(3)}\right)
 \equiv\left(-\sigma^2, i\sigma^3, -i\sigma^1\right)$ 
 and $V_0$ and $\gef{}$ are 
 the vacuum energy and 
 the effective Yukawa coupling 
\begin{equation}
 \gef{}\equiv\frac{g}{\sqrt{2}}\int^{\pi R}_{-\pi R}{\rm d}y \, 
 \cos\left(\frac{g}{\Lm}\Acl(y)\right) \bR{1}(y)\f{1}{0}(y)\f{2}{0}
(y)
 =\frac{g}{\sqrt{2}}\frac{C_{0}^{2}}{\CR{1}}(1-k^{2}).
 \label{geff}
\end{equation}

The nonvanishing mass term for $a_{{\rm R},1}$ shows a mass splitting 
associated to the SUSY breaking due to the coexistence of BPS 
and anti-BPS walls. 
The amount of this mass term can be related to the Yukawa coupling 
$g_{\rm eff}$ by means of low energy theorem. 
This fact provides a powerful method to evaluate the mass splitting 
between superpartners by evaluating the overlap of mode functions 
with the NG fermion \cite{MSSS}--\cite{MSSStw}. 

%%%%%%%%%%%%%%%%%%%%%%%%%%%%%%%%%%%%%%%

%\renewcommand{\thesubsection}{Acknowledgments}
%\subsection{}

One of the authors (N.S.) thanks the hospitality extended to him 
at the Corfu conference. 
This work is supported in part by Grant-in-Aid for Scientific 
Research from the Ministry of Education, Culture, Sports, 
Science and 
Technology, Japan, priority area (\#707) ``Supersymmetry and 
unified theory
of elementary particles" and No.13640269. 
N.M.,Y.S.~and R.S.~are supported 
by the Japan Society for the Promotion of Science for Young 
Scientists 
(No.08557, No.10113 and No.6665).

%%%%%%%%%%%%%%%%%%%%%%%%%%%%%%%%%%%%%%%%%%%%%%%%%%
%%%%%%%%%%%%%%%%%%%%%%%

\end{document}